\documentclass{article}

\usepackage{arxiv}

\usepackage[utf8]{inputenc} 
\usepackage[T1]{fontenc}    
\usepackage{hyperref}       
\usepackage{url}            
\usepackage{booktabs}       
\usepackage{amsfonts}       
\usepackage{nicefrac}       
\usepackage{microtype}      
\usepackage{lipsum}         
\usepackage{graphicx}
\usepackage{natbib}
\usepackage{doi}

\usepackage{tikz}
\usetikzlibrary{arrows}
\usepackage{array}
\usepackage{amsmath}
\usepackage{amssymb}
\usepackage{mathtools}
\usepackage{amsthm}
\usepackage[capitalise, noabbrev]{cleveref}
\PassOptionsToPackage{table}{xcolor}
\usepackage{colortbl}
\usepackage{pifont}
\usepackage{eurosym}
\usepackage{caption}
\usepackage{subcaption}

\newcommand*\diff{\mathop{}\!\mathrm{d}}

\DeclareMathOperator*{\argmin}{arg\,min}
\newcommand{\indep}{\perp \!\!\! \perp}

\theoremstyle{plain}
\newtheorem{theorem}{Theorem}[section]

\theoremstyle{definition}

\newtheorem{assumption}[theorem]{Assumption}
\theoremstyle{remark}

\newcommand{\PreserveBackslash}[1]{\let\temp=\\#1\let\\=\temp}
\newcolumntype{C}[1]{>{\PreserveBackslash\centering}p{#1}}
\newcolumntype{R}[1]{>{\PreserveBackslash\raggedleft}p{#1}}
\newcolumntype{L}[1]{>{\PreserveBackslash\raggedright}p{#1}}

\newcommand{\cmark}{\textcolor{black!75}{\ding{51}}}%
\newcommand{\xmark}{\textcolor{black!25}{\ding{55}}}%

\newcommand{\white}{\textcolor{white!100}{0}}%

\title{Prescriptive maintenance with causal machine learning}

\author{
    \href{https://orcid.org/0000-0001-7202-1589}{\includegraphics[scale=0.06]{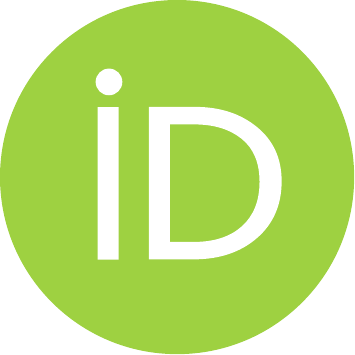}\hspace{1mm}Toon Vanderschueren\thanks{Corresponding author}}\\
	KU Leuven, University of Antwerp\\
	\href{mailto:toon.vanderschueren@kuleuven.be}{\texttt{toon.vanderschueren@kuleuven.be}} \\
	\And
	Robert Boute \\
	KU Leuven, Vlerick Business School, Flanders Make \\
	\href{mailto:robert.boute@kuleuven.be}{\texttt{robert.boute@kuleuven.be}} \\
	\And
	Tim Verdonck \\
	University of Antwerp, KU Leuven \\
	\href{mailto:tim.verdonck@uantwerpen.be}{\texttt{tim.verdonck@uantwerpen.be}} \\
	\And
	Bart Baesens \\
	KU Leuven, University of Southampton \\
	\href{mailto:bart.baesens@kuleuven.be}{\texttt{bart.baesens@kuleuven.be}} \\
	\And
	Wouter Verbeke \\
	KU Leuven \\
	\href{mailto:wouter.verbeke@kuleuven.be}{\texttt{wouter.verbeke@kuleuven.be}} \\
}

\renewcommand{\shorttitle}{\textit{arXiv} Template}

\hypersetup{
pdftitle={Prescriptive maintenance with causal machine learning},
pdfsubject={},
pdfauthor={Toon Vanderschueren, Robert Boute, Tim Verdonck, Bart Baesens, Wouter Verbeke},
pdfkeywords={Maintenance, Imperfect maintenance, Causal inference, Individual treatment effects, Machine learning},
}

\begin{document}
\maketitle

\begin{abstract}
    Machine maintenance is a challenging operational problem, where the goal is to plan sufficient preventive maintenance to avoid machine failures and overhauls. Maintenance is often imperfect in reality and does not make the asset as good as new. Although a variety of imperfect maintenance policies have been proposed in the literature, these rely on strong assumptions regarding the effect of maintenance on the machine's condition, assuming the effect is (1) deterministic or governed by a known probability distribution, and (2) machine-independent. This work proposes to relax both assumptions by learning the effect of maintenance conditional on a machine's characteristics from observational data on similar machines using existing methodologies for causal inference. By predicting the maintenance effect, we can estimate the number of overhauls and failures for different levels of maintenance and, consequently, optimize the preventive maintenance frequency to minimize the total estimated cost. We validate our proposed approach using real-life data on more than 4,000 maintenance contracts from an industrial partner. Empirical results show that our novel, causal approach accurately predicts the maintenance effect and results in individualized maintenance schedules that are more accurate and cost-effective than supervised or non-individualized approaches.
\end{abstract}


\section{Introduction}\label{sec:Introduction}

Machine maintenance constitutes an intricate operational problem. The challenge is to avoid machine failures and costly overhauls, while simultaneously minimizing the cost of preventive maintenance (PM). Moreover, maintenance is often imperfect in practice since it does not restore the machine to a state as good as new. In fact, a broad spectrum of maintenance effects have been studied in the literature, ranging from perfect maintenance, which restores the system to a state as good as new, to worst maintenance, where maintenance causes the machine to fail \citep{pham1996imperfect}. 

Existing approaches in imperfect maintenance rely on strong assumptions regarding the effect of PM. First, the effect is modelled as either deterministic or stochastic assuming a certain probability distribution. These assumed effects, however, might not correspond to the actual effect. Second, the effect is typically assumed to be machine-independent, i.e., identical for all machines. In reality, the effect of the same type of PM intervention could be very different for different machines. For example, changing a gear would likely have a different impact on a brand new machine compared to the exact same maintenance intervention on an old, worn down machine.

This work relaxes both assumptions by proposing a completely data-driven maintenance policy that learns the effect of maintenance conditional on a machine's characteristics. The benefit of this approach is that it allows (1) to flexibly learn the maintenance effects from observational data (instead of assuming a certain deterministic or stochastic effect based on expertise), and (2) to design a machine-specific PM schedule based on these learned effects. 

These benefits are achieved by framing maintenance as a problem of causal inference. We argue that the challenge in maintenance is that, for each specific machine, we only observe one outcome for the maintenance frequency that was administered in practice. We never observe the counterfactual outcomes -- what would have happened if that machine received more or less maintenance in the past. Therefore, we never know whether the optimal maintenance frequency was prescribed. This is exactly the aim of causal inference, i.e. to predict each individual machine's potential outcomes in terms of failures and overhauls for different levels of PM. By learning a model that predicts the number of overhauls and failures given the PM frequency, we can optimize the PM schedule to minimize the total estimated cost. Essentially, we propose using observational data to learn a machine-specific digital twin for maintenance that predicts what would happen if a machine is prescribed a certain maintenance schedule.

This work contributes by proposing a novel prescriptive framework for maintenance that prescribes maintenance based on the estimated effect of PM on the machine's number of overhauls and failures. To this aim, we frame maintenance as a problem of causal inference. Consequently, we leverage state-of-the-art machine learning methods for causal inference that learn models to estimate a machine's potential outcomes for different PM frequencies from observational data.
Moreover, we formulate a prescriptive policy that uses the potential outcomes to decide on the optimal PM frequency so as to minimize the total cost of failures and interventions. Empirically, we contribute by demonstrating the excellent use of the presented prescriptive framework on a dataset consisting of more than 4,000 maintenance contracts of industrial equipment provided by an industrial partner.

\section{Related work}\label{sec:Related_work}

Machine maintenance has been studied extensively in operations research, with a wide variety of proposed maintenance policies \citep{wang2002survey, ding2015maintenance, de2020review}. Although most existing work assumes that maintenance restores the system to a state that is as good as new, maintenance is typically imperfect in reality. In fact, different maintenance effects have been studied in the literature, ranging from maintenance that restores the system to a perfect state to maintenance that makes the system's state worse \citep{pham1996imperfect}. Consequently, developing maintenance policies that incorporate imperfect maintenance is an important research problem.

\subsection{Imperfect maintenance}
Existing work models the effect of imperfect maintenance as either stochastic (based on a known probability distribution) or deterministic \citep{pham1996imperfect, chukova2004warranty}. Stochastic effects include the $(p,q)$ rule, where maintenance is perfect with probability $p$ and minimal with probability $q=1-p$ \citep{nakagawa1979imperfect, nakagawa1979optimum, brown1983imperfect}, as well as its age-dependent variant $(p(t), q(t))$ \citep{block1985age}. Other work assumes a deterministic effect. Improvement factor models assume that maintenance decreases the system's failure rate by a deterministic improvement factor  \citep{malik1979reliable}. Similarly, in virtual age models, imperfect maintenance decreases the system's age or its failure intensity with a deterministic factor $q$ where $0 < q < 1$ \citep{kijima1989some, tanwar2014imperfect}.

Conversely, instead of making assumptions regarding the effect of maintenance, our work proposes to learn the effect of PM from data. Data-driven approaches have recently gained importance in the maintenance literature \citep{bousdekis2021review}. Condition-based maintenance is a recent paradigm where maintenance is optimized based on the machine's state or its characteristics \citep{gits1992design, alaswad2017review}. Especially relevant to our work are recent, predictive maintenance approaches that learn a predictive model from data to decide on the appropriate maintenance interventions \citep{swanson2001linking, carvalho2019systematic}. 

Similar to the general literature on imperfect maintenance, existing condition-based approaches that do consider imperfect maintenance assume either a deterministic or stochastic maintenance effect. There exist three broad categories of condition-based approaches that account for imperfect maintenance \citep{alaswad2017review}. A first category considers minimal maintenance with a \textit{deterministic} effect, in which a system has several deterioration stages and imperfect maintenance returns the system to the previous stage. A second category considers \textit{stochastic} effects where the maintenance effect is governed by an assumed probability distribution. Finally, in \textit{improvement factor} models, imperfect maintenance decreases the system's hazard rate with a (deterministic) factor between zero and one. To the best of our knowledge, no existing condition-based approaches aim to \textit{learn} the effect of maintenance from data. 

Finally, this work focuses on a provider of full-service contracts

\subsection{Prescriptive analytics and causal inference}
Instead of assuming a certain PM effect, this work uses data-driven models to learn the effect of maintenance using techniques from causal inference. Causal inference aims to estimate the effect of a certain cause from data, e.g., the number of failures resulting from a certain maintenance frequency compared to not applying any maintenance. Ideally, estimating maintenance effects would be done by conducting a randomized controlled trial: assigning different levels of maintenance to a collection of (similar) machines and comparing the outcomes \citep{rubin1974estimating}. However, in practice, this approach can be prohibitively expensive or even unfeasible. In maintenance specifically, it would be challenging to randomly assign various levels of PM to different machines. Therefore, we need to rely on historical, observational data of machines and their maintenance.

The challenge of working with observational data is that this data is biased due to existing maintenance policies that were applied \citep{rubin1974estimating}. For example, as a result of an existing policy, machines more prone to failure might have been more likely to receive maintenance in practice. This phenomenon, called selection bias or confounding bias, can result in biased estimates of the counterfactual outcomes if ignored. Therefore, specialized tools have been developed in the causal inference literature to tackle exactly this problem and learn causal effects from observational data, i.e., in the presence of selection bias \citep{yao2021survey}. Specifically, our work is related to learning potential outcomes for continuous-valued interventions \citep{imbens2000role, hirano2004propensity, imai2004causal, schwab2020learning, bica2020estimating}, e.g., the number of PM interventions per running period.

Causal inference has been applied to a variety of applications, such as personalized medicine \citep{berrevoets2020organite}, economic policy design \citep{athey2021policy}, or marketing \citep{varian2016causal, devriendt2018literature}. Moreover, it is related to prescriptive analytics \citep{verbeke2020foundations, verbeke2022or}, which has recently gained importance in operations research \citep{bertsimas2019optimal, bertsimas2020predictive}. In this work, causal inference is used to predict a machine's potential outcomes for different levels of maintenance and decide upon a personalized maintenance schedule. To the best of our knowledge, this is the first application of causal inference for maintenance optimization.

\section{Problem overview}\label{sec:Problem_overview}

This work aims to solve the problem faced by a provider of full-service maintenance contracts. The service provider is responsible for maintaining the client's asset at a predetermined price \citep{deprez2021pricing}. Therefore, for each contract, the service provider needs to decide on a usage-based PM schedule, prior to contract start, based on information such as the type of machine it concerns and/or the machine's age at contract start.

In this work, we assume the service provider conducts a single type of PM intervention and needs to decide on the frequency of these interventions. Planned PM aims to prevent two types of events. The first, overhauls, are unplanned, comprehensive maintenance interventions during which large parts of the machinery need to be replaced. From the viewpoint of the full-service maintenance provider, these are the most costly type of event. The second, machine failures, result in an urgent need for maintenance as the machine stops running until corrective maintenance occurs. This again incurs a cost to the service provider that is smaller than the cost of an overhaul, but larger than the cost of PM.

The overall goal is to find each contract's optimal PM frequency that minimizes the combined cost of planned PM, overhauls and failures, from the perspective of the service provider. Although planning more PM interventions is likely to result in less overhauls and failures, it also comes at an increased maintenance cost. This means the PM frequency is a trade-off between costs resulting from planned PM on the one hand and costs resulting from overhauls and failures on the other hand. Due to heterogeneity in the contracts, maintenance might need to be planned more frequently for some machines. Therefore, it is important to consider the contract's characteristics when deciding on the PM frequency. To this aim, the service provider has access to information on past contracts including how often maintenance was applied as well as the number of overhauls and failures that were observed.

Formally, each contract is defined as a tuple $(\mathbf{X}, T, O, F)$. $\mathbf{X} \in \mathcal{X} \subset \mathbb{R}^d$ denotes the characteristics of the machine and contract. The treatment, the PM frequency or the number of preventive maintenance interventions that will be applied per running period, is denoted as $T \in \mathcal{T} \subset \mathbb{R}^+$. $O \in \mathcal{O} \subset \mathbb{R}^+$ and $F \in \mathcal{F} \subset \mathbb{R}^+$ are the contract's number of overhauls and failures per running period. We adopt the Rubin--Neyman potential outcomes framework \citep{rubin2004direct, rubin2005causal} and denote the overhauls $O$ and failures $F$ per running period given maintenance frequency $t$ as $O(t)$ and $F(t)$.

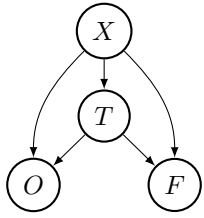
\begin{figure}[!hbt]
\centering
\begin{tikzpicture}%
    [vertex/.style={circle,draw=black,fill=white},
    >=latex
    ]
    
    \tikzstyle{round}=[circle,thick,auto,draw]
    
    \matrix[column sep=0.2cm, row sep=0.2cm,column 6/.style={anchor=base west}] {
     & \node[round] (X) {$X$}; & \\
     & & \\
     & \node[round] (T) {$T$}; & \\
    \node[round] (O) {$O$}; & & \node[round] (F) {$F$};\\
    };
    \draw[->]
        (X) edge (T)
        (X) edge [out=225, in=90] (O)
        (X) edge [out=315, in=90] (F)
        (T) edge (O)
        (T) edge (F)
        ;
\end{tikzpicture}%
\caption{\textbf{Causal diagram depicting the relations between the different variables.} $X$: Machine and contract characteristics, $T$: Preventive maintenance, $O$: Overhauls, and $F$: Failures.}
\label{fig:dag}
\end{figure}

The objective is to decide on the optimal maintenance frequency $t_i^*$ that minimizes the total cost per running period. We assume a usage-based maintenance cost similar to \cite{faccio2014industrial}. A machine $i$'s cost per running period given PM frequency consists of the combined costs of PM, overhauls and failures:
\begin{equation}\label{eq:total_cost}
    c_i(t_i) = 
    \underbrace{c_t \, t_i \,  \vphantom{c_f}}_{\text{PM}} + 
    \underbrace{c_o \, o_i \, \vphantom{c_f}}_{\text{Overhauls}} + 
    \underbrace{c_f \, f_i}_{\text{Failures}}.
\end{equation}
Here, we assume that the individual costs of preventive maintenance, overhauls and failures ($c_t, c_o, c_f \in \mathbb{R}^+$) are deterministic and known.

To assist the full service-provider's decision-making, a data set is available with information on $n$ past contracts $\mathcal{D} = \left\{(\mathbf{x}_i, t_i, o_i, f_i)\right\}^{n}_{i=1}$. For each of the past contracts, only one potential outcome was observed for $O$ and $F$: $o_i(t)$ and $f_i(t)$. The other, counterfactual outcomes are never observed. This is known as the fundamental problem of causal inference \citep{holland1986statistics}. The challenge in causal inference is to predict, for a new contract, all potential outcomes by learning from this historical data. 

Because past decisions regarding the PM frequency were made according to an unknown existing policy, there is selection bias in the data. This means that contracts that were likely to receive relatively little PM are different from machines that were likely to receive relatively much PM. For example, the service provider might have known from experience that a certain type of machine would be likely to fail often when not receiving frequent PM and, because of this, prescribed more maintenance to those machines in the past. Therefore, learning a predictive model for estimating potential outcomes from observational data needs to adjust for selection bias in this data to obtain unbiased estimates.

\section{Methodology}\label{sec:Methodology}

\begin{figure*}
    \centering
    \includegraphics[width=\linewidth]{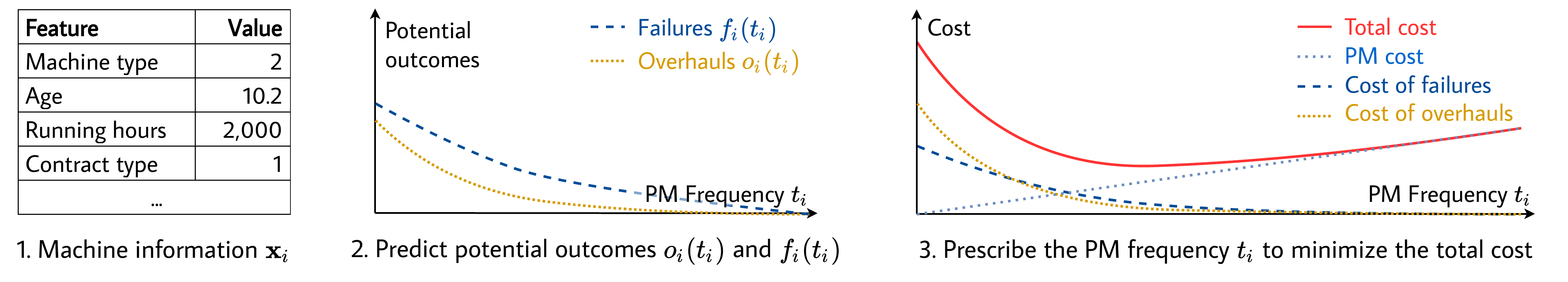}
    \caption{\textbf{Methodology overview.} We present a high-level overview of our methodology. Machine characteristics $\mathbf{x}_i$ are used to predict the potential outcomes in terms of overhauls $o_i(t)$ and failures $f_i(t)$. Based on these estimates, the total cost for different levels of maintenance can then be estimated. Finally, the PM frequency is chosen to minimize the total expected cost.}
    \label{fig:methodology_overview}
\end{figure*}

Our methodology consists of a predict-then-optimize framework, see \cref{fig:methodology_overview} for a high-level overview.
To estimate each contract's cost for a certain PM frequency $c_i(t_i)$, each machine's potential outcomes need to be estimated, i.e., its number of overhauls $o_i(t)$ and failures $f_i(t)$ for a PM frequency $t_i$, given its characteristics $\mathbf{x}_i$. Therefore, the first step is to learn a machine learning model for estimating potential outcomes from historical, observational data on similar full-service contracts $\mathcal{D}$. In a second phase, these estimated outcomes can be used to optimize the PM frequency and resulting total cost.

The rest of this section is organized as follows. First, estimating potential outcomes from observational data requires two standard assumptions. These are put forward in \cref{ssec:assumptions}. Second, we estimate the potential outcomes by learning a predictive model from observational data. For this, we use a state-of-the-art methodology called SCIGAN \citep{bica2020estimating}, which is described in \cref{ssec:predict}. Third, in \cref{ssec:optimize}, these predictions are used to assign each machine's optimal PM frequency that minimizes the total estimated cost.

\subsection{Assumptions}\label{ssec:assumptions}
The challenge in estimating potential outcomes from observational data is dealing with selection bias. Learning unbiased estimates of the potential outcomes from observational data requires making three standard assumptions: consistency, overlap and unconfoundedness \citep{imbens2000role, bica2020estimating}. Given these assumptions, adjusting for machine characteristics $\mathbf{x}_i$ allows to account for selection bias in observational data and obtain unbiased estimates. The first assumption is consistency, i.e., a machine's potential outcome given observed treatment $t$ is the observed outcome.
\begin{assumption}{\textbf{Consistency.}}\label{ass:consistency}
    $Y = Y(t)$ for all $t \in T$.
\end{assumption}
The second, overlap or positivity, ensures that each possible contract $\mathbf{x}_i$ has a non-zero probability of receiving each frequency of PM interventions $t_i$.
\begin{assumption}{\textbf{Overlap.}}\label{ass:overlap}
    For all $\mathbf{x} \in \mathcal{X}$ with $p(\mathbf{x}>0)$ and $\, t \in \mathcal{T}$: 
    $0 < p(t|\mathbf{x}) < 1$.
\end{assumption}
\noindent
The third, unconfoundedness or no hidden confounders, ensures that there are no unobserved variables influencing both the treatment assignment $T$ and a potential outcome $O(t)$ or $F(t)$.
\begin{assumption}{\textbf{Unconfoundedness.}}\label{ass:unconfoundedness}
    Potential outcomes $O(t)$ and $F(t)$ are independent of the PM frequency $T$ conditional on machine characteristics $X$:
    $\{O(t), F(t) \; | \; t\in \mathcal{T}\} \indep T \; | \; \mathbf{X}$.
\end{assumption}
\noindent

\subsection{Predicting preventive maintenance effects}\label{ssec:predict}

First, we need to predict each machine's potential outcomes $o_i(t)$ and $f_i(t)$ given a PM frequency $t_i$ based on characteristics $\mathbf{x}_i$. Therefore, we aim to find models $g_o: \mathcal{X}\times\mathcal{T} \to \mathcal{O}$ and $g_f: \mathcal{X}\times\mathcal{T} \to \mathcal{F}$ defined by parameters $\theta_o, \theta_g \in \Theta$ and obtain unbiased estimates of the potential outcomes $g_o(t, \mathbf{x}) = \mathbb{E}\left[O(t) \; | \; \mathbf{X} = \mathbf{x}\right] \; \text{ and } \; g_f(t, \mathbf{x}) = \mathbb{E}\left[F(t) \; | \; \mathbf{X} = \mathbf{x}\right]$.

In this work, $g_o$ and $g_f$ are learned using SCIGAN, a recently proposed machine learning approach for predicting potential outcomes given a continuously-valued treatments \citep{bica2020estimating}. SCIGAN achieved state-of-the-art performance across a variety of settings. $g$ is learned in two steps. First, a generative adversarial network (GAN) is trained to model the distribution of the potential outcomes: the generator is trained to generate counterfactual contracts that cannot be distinguished from factual, observed contracts by the discriminator. In a second phase, the GAN is used to augment the observed training data with generated counterfactual samples. This way, the augmented data set contains all potential outcomes, including both the factual outcomes and the generated, counterfactual outcomes. Because of this, selection bias is no longer a problem and, using this augmented data set, a predictive model $g_\theta$ can be trained to predict the potential outcomes in a supervised manner. For this, we use a neural network. More specificallly, we use a multilayer perceptron (MLP).

\subsection{Optimizing the maintenance cost}\label{ssec:optimize}

The optimal PM frequency is a trade-off between costs resulting from planned PM on the one hand and costs resulting from overhauls and failures on the other hand. However, using the potential outcomes $o_i(t_i)$ and $f_i(t_i)$, it can be seen that the overhauls and failures can be written as functions of the PM frequency $t_i$. Therefore, the predicted potential outcomes can be used to directly estimate the costs incurred at different PM frequencies. This is achieved by rewriting all terms in \cref{eq:total_cost} (PM, overhauls and failures) as a function of the PM frequency $t_i$:
\begin{equation}\label{eq:total_cost_t}
    c_i(t_i) = c_t \, t_i + c_o \, o_i(t_i) + c_f \, f_i(t_i).
\end{equation}
Each machine's optimal PM frequency $t_i^*$ is found as the level that minimizes the expected cost: $t_i^* = \argmin c_i(t)$. To account for heterogeneity in the contracts, this optimal PM frequency is optimized for each specific machine.

\section{Results}\label{sec:Results}

We validate our methodology empirically using real-world data on full-service maintenance contracts. The goal is to decide on the optimal PM frequency, prior to the contract start, to minimize the total cost resulting from preventive maintenance, overhauls and failures.

\subsection{Data}\label{ssec:Data}

Our data set contains more than 4,000 full-service maintenance contracts. For each contract $i$, we have information $\mathbf{x}_i$ on the machine, the contract itself, and maintenance-related events (see \cref{tab:data_overview}). Events are presented per running period, which is a set number of running hours. For reasons of confidentiality, the exact number of running hours per period is not presented. Costs are averaged over all events and re-scaled for reasons of confidentiality.

The data is preprocessed as follows. Categorical variables are encoded with dummies and $\mathbf{x}_i$ is standardized. The PM frequency, overhauls and failures that occurred throughout the contract are converted to the number of events per running period. Even though a contract's exact number of running hours is not known in advance, an estimate is typically available. 

\begin{table}[]
    \centering
    \begin{tabular}{L{150pt}R{100pt}}
    \toprule
        \textbf{Variable}    & \textbf{Domain} \\
    \midrule
        \rowcolor{gray!10} \multicolumn{2}{c}{\textbf{Machine information}}  \\
        Type                            & $\{1, \dots, 7\}$ \\
        Age at contract start           & $[0, 39]$\\
        Running hours at contract start & $[2500, 110000]$\\
        Running hours during contract   & $[0, 186000]$\\
        Average running hours per year  & $[300, 8500]$\\
    \midrule
        \rowcolor{gray!10} \multicolumn{2}{c}{\textbf{Contract information}} \\
        Type                & $\{1, 2\}$    \\
        Duration (days)     & $[180, 5850]$ \\
    \midrule
        \rowcolor{gray!10} \multicolumn{2}{c}{\textbf{Preventive maintenance per running period}} \\
        PM frequency        & $[0, 20]$ \\
    \midrule
        \rowcolor{gray!10} \multicolumn{2}{c}{\textbf{Outcomes per running period}} \\
        Number of overhauls & $[0, 128]$ \\
        Number of failures  & $[0, 185]$ \\
    \midrule
        \rowcolor{gray!10} \multicolumn{2}{c}{\textbf{Average costs (in \euro)}} \\
        Preventive maintenance  &  $73$ \\  
        Overhaul                & $207$ \\  
        Failure                 & $104$ \\  
    \bottomrule \\
    \end{tabular}
    \caption{\textbf{Data overview.} Overview of the available contract information on machine and contract characteristics, preventive maintenance interventions, overhauls, and failures.}
    \label{tab:data_overview}
\end{table}

\subsection{Semi-synthetic setup}\label{ssec:Semi_synthetic_data}
A good estimator should accurately predict both the observed outcome, the number of failures that did occur at maintenance frequency $t_i$, as well as the unobserved outcomes, the number of failures if the machine had received less or more maintenance. In practice however, not all potential outcomes are observed, which makes evaluation of causal models hard. Because of this, we rely on semi-synthetic data to evaluate our model. This approach is commonly used in both causal inference \citep[see e.g.,][]{berrevoets2020organite} and maintenance \citep[e.g., ][]{deprez2021pricing}.

Potential outcomes $o_i(t)$ and $f_i(t)$ are generated based on the observed characteristics $\mathbf{x}_i$. For the overhauls, we have:
\begin{equation}
    o_i(t) = 7 \, \sigma\Bigg(
    {\underbrace{\mathbf{v}_o^\intercal \mathbf{x}_i \vphantom{\frac{1}{10}}}_{\text{Base rate}}} - \, 
    {\underbrace{\frac{1}{10} \, \sigma\left(\mathbf{w}_o^\intercal \mathbf{x}_i \right)}_{\text{PM effect}}} t \, +
    {\underbrace{\epsilon_o \vphantom{\frac{1}{10}}}_{\text{Noise}}}
    \Bigg)    
\end{equation}
where $\mathbf{v}_o, \mathbf{w}_o \sim \mathcal{U}\left((0, 1)^{d\times1}\right)$ and $\epsilon_o \sim \mathcal{N}(0, 1)$. The 7 rescales the average number of overhauls to roughly same number in the original data. For failures, we similarly have:
\begin{equation}
    f_i(t) = 9 \, \sigma\bigg(\mathbf{v}_f^\intercal \mathbf{x}_i - \frac{1}{10} \, \sigma\left(\mathbf{w}_f^\intercal \mathbf{x}_i\right) t + \epsilon_f\bigg)    
\end{equation}
with $\mathbf{v}_f, \mathbf{w}_f \sim \mathcal{U}\left((0, 1)^{d\times1}\right)$ and $\epsilon_f \sim \mathcal{N}(0, 1)$.

\begin{figure}
\centering
\begin{subfigure}{.5\linewidth}
  \centering
  \includegraphics[height=100pt]{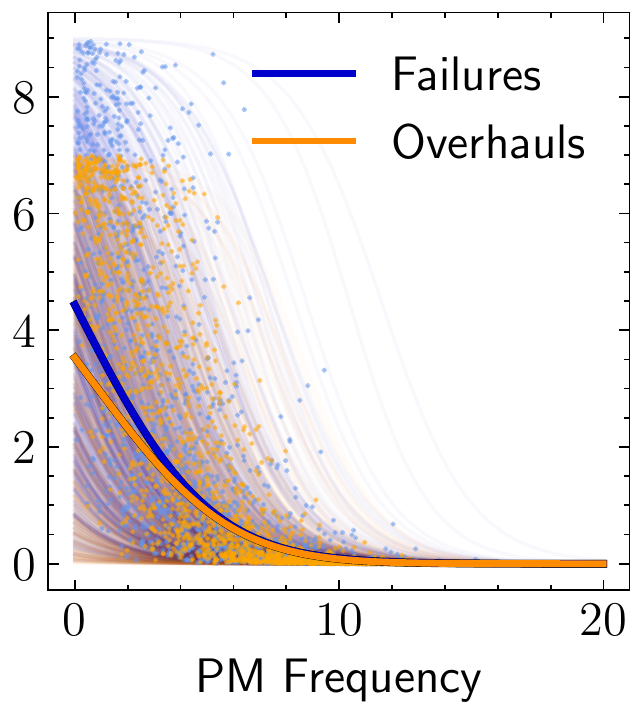}
  \caption{Potential outcomes}
  \label{sfig:data}
\end{subfigure}%
\begin{subfigure}{.5\linewidth}
  \centering
  \includegraphics[height=100pt]{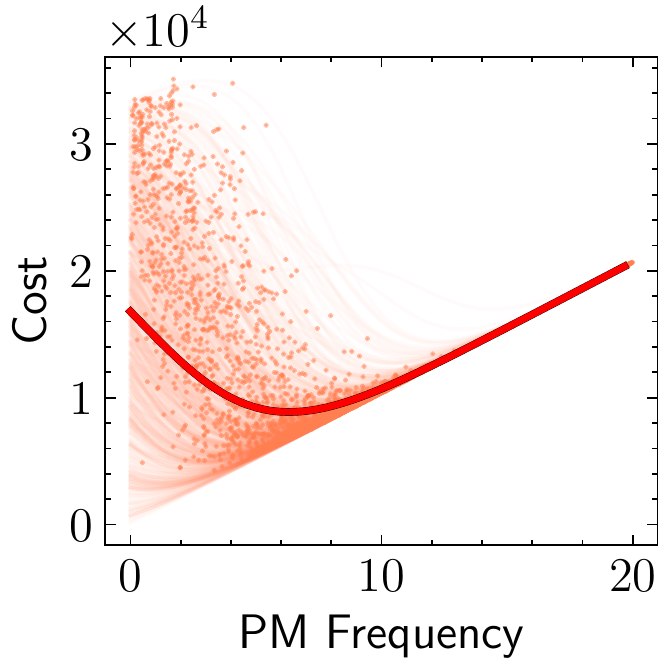}
  \caption{Costs}
  \label{sfig:costs}
\end{subfigure}
\caption{\textbf{Semi-synthetic data.} We show the observed outcomes in the training and validation set with dots and potential outcomes in the test set with a line. The average potential outcomes and cost are shown with a bold line.}
\label{fig:semi_synthetic_data}
\end{figure}

Using the semi-synthetic setup, the test set contains the potential outcomes for all possible values of $t_i \in \mathcal{T}$ using these equations. Conversely, the training and validation sets include only one observed outcome for one observed $t_i$. The training, validation and test sets respectively consist out of 50\%, 25\% and 25\% of the data. Hyperparameter optimization is based on the mean squared error on the observed outcomes in the validation set. An illustration of a generated data set is shown in \cref{fig:semi_synthetic_data}.

\begin{figure*}
    \centering
    \begin{subfigure}[b]{0.79\textwidth}
         \centering
         \includegraphics[height=85pt]{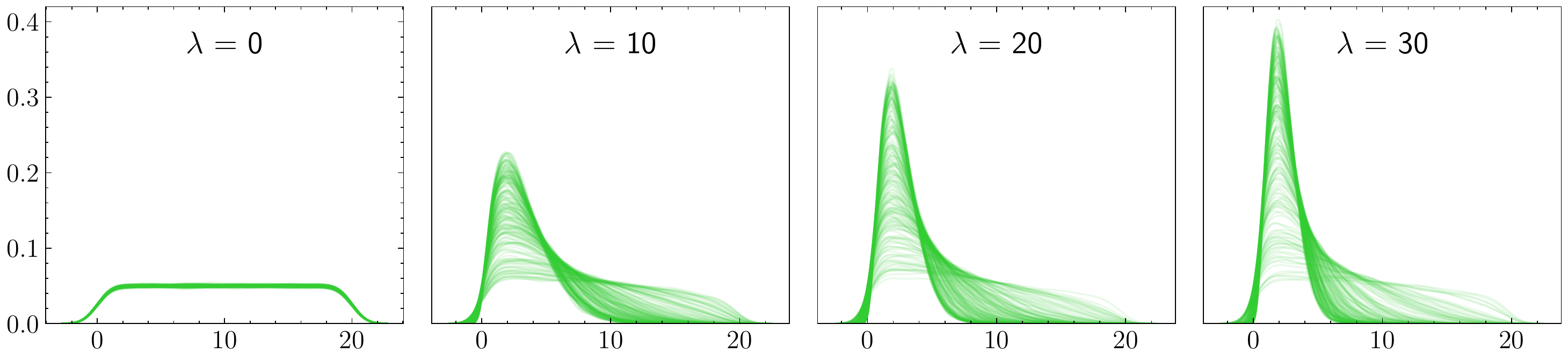}
         \caption{Effect of $\lambda$ on the individual machines' maintenance distributions}
         \label{sfig:lambda_effect}
     \end{subfigure}
     \hfill
     \begin{subfigure}[b]{0.2\textwidth}
         \centering
         \includegraphics[height=85pt]{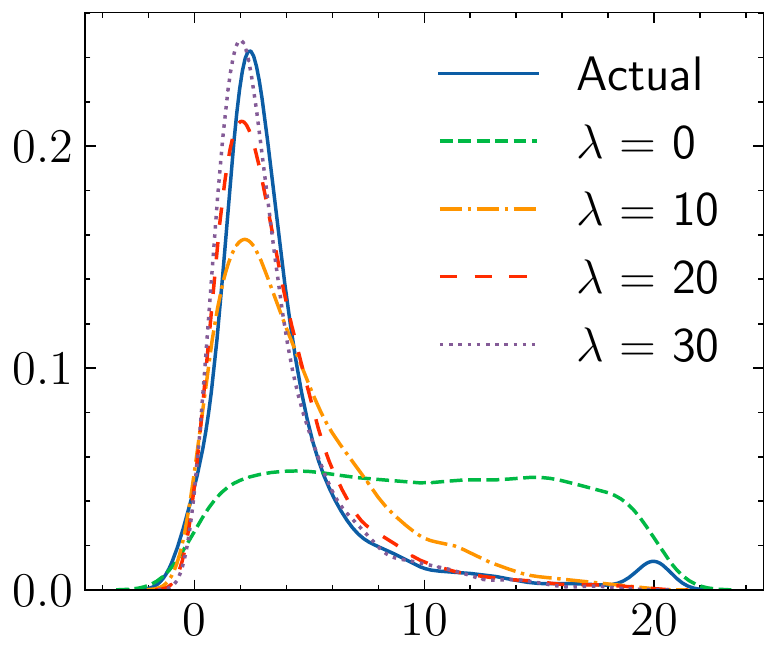}
         \caption{PM frequencies}
         \label{sfig:pm_distribution}
     \end{subfigure}
    \caption{\textbf{Simulating selection bias.}
    (\ref{sfig:lambda_effect}) We show the distributions that govern the PM frequency for different machines. As these distributions depend on the machine's characteristics, certain machines will more frequently have more maintenance, resulting in selection bias. Moreover, higher values of $\lambda$ imply more diversity in the distributions and, consequently, more selection bias.
    (\ref{sfig:pm_distribution}) We show how the PM frequency is distributed among the different machines in reality and as a result of different values of $\lambda$. Larger values of $\lambda$ result in more selection bias with a value of $30$ resulting in a PM frequency distribution close to the original.}
    \label{fig:beta_distribution}
\end{figure*}

On the one hand, we evaluate the prescribed maintenance frequencies using the maintenance frequency $t_i$ that was observed in practice for the observed outcomes in the training and validation set. On the other hand, we want to evaluate our policy for different levels of selection bias. For this, we control the level of selection bias in the semi-synthetic data using an approach similar to \cite{bica2020estimating}. Selection bias is simulated by assigning PM frequencies from a beta distribution as follows:
\begin{equation}
    t_i \sim 20 \, \text{Beta}\left(1 + \frac{\lambda \delta_i}{10}, 1 + \lambda \delta_i\right)
\label{eq:sim_selection_bias_lambda}
\end{equation}
where $\delta_i = \sigma(\textbf{w}_b \mathbf{x}_i)$ with $\textbf{w}_b \sim \mathcal{U}\left((0, 1)^{d\times1}\right)$. $\delta_i$ ensure that treatment assignment is based on observed features $\mathbf{x}_i$. This way, $\lambda$ controls the level of selection bias. $\lambda = 0$ results in $\text{Beta}(1, 1)$ or the uniform distribution, which implies random maintenance assignment. Higher values of $\lambda$ imply more selection bias with $\lambda=30$ resulting in a maintenance distribution similar to the observed distribution. An illustration of the observed distribution and generated distributions for different values of $\lambda$ is shown in \cref{fig:beta_distribution}. 

\subsection{Evaluation}\label{ssec:Evaluation}

Evaluation is done using three different metrics. First, we evaluate the ability of the machine learning model to accurately predict a contract's potential outcomes. This is measured using the mean integrated square error (MISE) \citep{silva2016observational, schwab2020learning}:
\begin{equation}
    \text{MISE} = \frac{1}{n} \sum_{i=1}^n \int_0^m \left(y_i(t) - \hat{y}_i(t) \right)^2 \,\diff t
    .
\end{equation}
Second, we want to evaluate the accuracy of the prescribed maintenance frequencies. To this end, we consider a variant of the policy error (PE) \citep{schwab2020learning} that compares the prescribed maintenance frequency with the ideal level:
\begin{equation}
    \text{PE} = \frac{1}{n} \sum_{i=1}^n \left(t^*_i - \hat{t}^*_i \right)^2
    .
\end{equation}
Third, we evaluate the prescribed maintenance frequency in terms of costs using the policy cost ratio (PCR) that compares the costs of the estimated optimal maintenance frequency with the ideal level:
\begin{equation}
    \text{PCR} = \frac{1}{n} \sum_{i=1}^n \frac{c_i(\hat{t}_i^*)}{c_i(t_i^*)}
    .
\end{equation}
For all metrics, a lower value indicates better performance with $0$ being the optimal value for MISE and PE and $1$ for PCR.

Our proposed maintenancy policy uses SCIGAN to learn the individual treatment effects (ITE) and will be referred to as SCIGAN--ITE. We benchmark this policy to two other policies (see \cref{tab:methodologies}). First, a policy based on a neural network (MLP) that learns $o_i$ and $f_i$ given $\mathbf{x}_i$ and $t_i$ in a completely supervised manner without adjusting for selection bias (MLP--ITE). This allows us to assess whether there is a benefit of adjusting for selection bias. Second, the average policy (SCIGAN--ATE) sets a single optimal $t^*$ for all contracts based on the average (instead of the individual) maintenance effect. This allows to validate the benefit of an individualized policy tailored towards each specific machine.

\begin{table}[]
    \centering
    \begin{tabular}{L{68pt}C{68pt}C{68pt}}
    \toprule
         \textbf{Methodology}    & \textbf{Selection bias?} & \textbf{Individualized?} \\
     \midrule
         SCIGAN--ITE  & \cmark    & \cmark          \\
     \midrule
         MLP--ITE     & \xmark    & \cmark          \\
         SCIGAN--ATE  & \cmark    & \xmark          \\
    \bottomrule \\
    \end{tabular}
    \caption{\textbf{Methodologies overview.} Our proposed, individual policy, SCIGAN--ITE, prescribes the PM frequency based on the individual treatment effect (ITE) estimated using SCIGAN. This proposed approach is analyzed using an ablation study and compared with two variants. The first, MLP--ITE, does not account for selection bias. The second, SCIGAN--ATE, is a general policy based on the average treatment effect (ATE) and is not individualized towards each individual machine.}
    \label{tab:methodologies}
\end{table}

\subsection{Empirical results}\label{ssec:Empirical_results}

In this section, we present the results of the semi-synthetic experiments based on more than 4,000 maintenance contracts, as put forward in \cref{ssec:Data,ssec:Semi_synthetic_data,ssec:Evaluation}. The goal is to answer two research questions. First, does an individualized approach outperform a general approach? Second, does a causal, prescriptive approach outperform a supervised, predictive approach? We aim to answer these for the observed maintenance frequency (\cref{sssec:observed_results}) and assess the different policies' sensitivity to varying levels of selection bias (\cref{sssec:selection_bias_results}).

\subsubsection{Results for the observed PM frequencies}
\label{sssec:observed_results}

\begin{figure}
    \centering
    \includegraphics[width=0.55\linewidth]{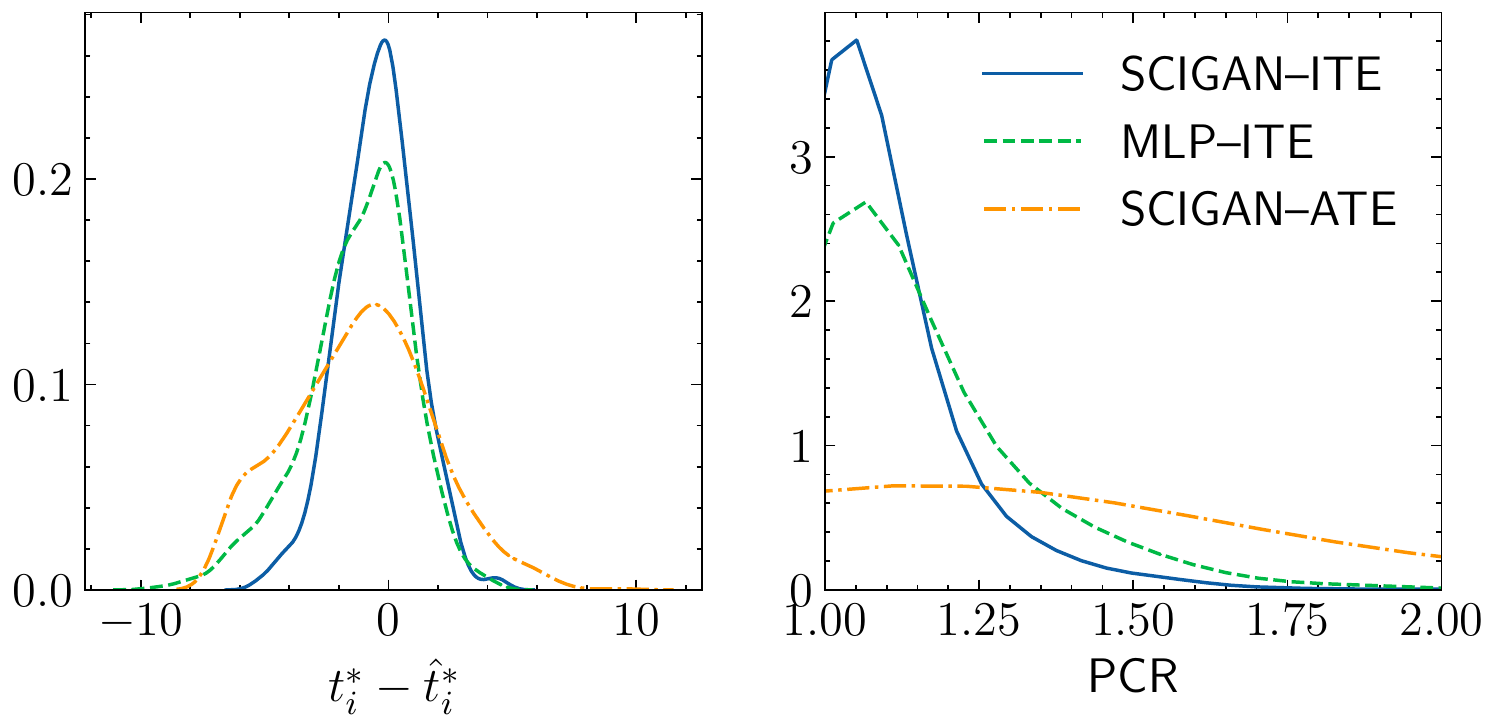}
    \caption{\textbf{Evaluating the policies' decisions.} We compare the accuracies and costs of each policy's prescribed decisions in terms of the difference between the prescribed and ideal maintenance level (left), as well as the policy cost ratio (right). Results are shown for one representative iteration.}
    \label{fig:decisions_comparison}
\end{figure}

We present the results for the different methodologies given the maintenance frequency $t_i$ that was observed in practice in \cref{tab:results} and \cref{fig:decisions_comparison}. For both failures and overhauls, SCIGAN more accurately predicts the potential outcomes in terms of MISE compared to MLP, the supervised approach. Moreover, the individualized, prescriptive approach (SCIGAN--ITE) most accurately prescribes the optimal PM frequency compared to the supervised (MLP--ITE) and non-individualized approach in terms of policy error. Finally, SCIGAN--ITE also results in lower costs compared to both MLP--ITE and SCIGAN--ATE. The improved performance of SCIGAN--ITE compared to MLP--ITE illustrates the importance of adjusting for selection bias when learning from observational data. Moreover, the relatively worse performance of the average approach, SCIGAN--ATE, indicates the benefit of an individualized, machine-dependent policy for imperfect maintenance that takes into account machine characteristics.

\begin{table*}
    \centering
    \begin{tabular}{L{50pt}C{65pt}C{65pt}}
    \toprule
     & \multicolumn{2}{c}{$\textbf{MISE}$} \\
     & \multicolumn{1}{c}{Overhauls} & \multicolumn{1}{c}{Failures} \\
    \midrule
        SCIGAN & $\mathbf{\white7.71 \ \pm 0.60}$ & $\mathbf{14.16 \ \pm 1.68}$ \\
        MLP    &               $10.25 \ \pm 1.33$ &           $18.27 \ \pm3.65$ \\
    \bottomrule
    \end{tabular}
    \hfill
    \begin{tabular}{L{70pt}C{65pt}C{65pt}}
    \toprule
     & \textbf{PE} & \textbf{PCR} \\
    \midrule
        SCIGAN--ITE & $\mathbf{2.40 \ \pm 0.46}$ & $\mathbf{1.07 \ \pm 0.01}$ \\
        MLP--ITE    &          $4.36 \ \pm 1.25$ &          $1.11 \ \pm 0.02$ \\
        SCIGAN--ATE &                $8.77 \ \pm 1.07$ &          $1.24 \ \pm 0.04$ \\
    \bottomrule
    \end{tabular}
    \caption{\textbf{Empirical evaluation.} We compare performance for the different policies over five runs. We evaluate each model's ability to predict the potential outcomes $o_i(t)$ and $f_i(t)$ (MISE), as well as each policy's ability to accurately prescribe the maintenance frequency (PE) and minimize costs (PCR). For all metrics, a lower value is better.}
    \label{tab:results}
\end{table*}

\subsubsection{Results for different levels of selection bias}\label{sssec:selection_bias_results}
We compare performance for the SCIGAN--ITE and MLP--ITE for different levels of selection bias in terms of $\lambda$ (see \cref{eq:sim_selection_bias_lambda}). SCIGAN achieves good predictive performance in terms of MISE for the entire range of operating conditions, ranging from randomized PM assignments ($\lambda = 0$) to realistic levels of selection bias ($\lambda = 30$). Conversely, the MLP, a completely supervised approach that does not adjust for selection bias, accurately predicts the potential outcomes when preventive maintenance is randomized ($\lambda=0$), but results in bad predictions for high levels of $\lambda$. Similarly, SCIGAN is robust towards higher levels of $\lambda$ in terms of decision-making, illustrated by stable values for PE and PCR across different levels of selection bias, whereas MLP results in less accurate and more costly decisions as bias increases.

\begin{figure*}[!h]
    \centering
    \includegraphics[width=\linewidth]{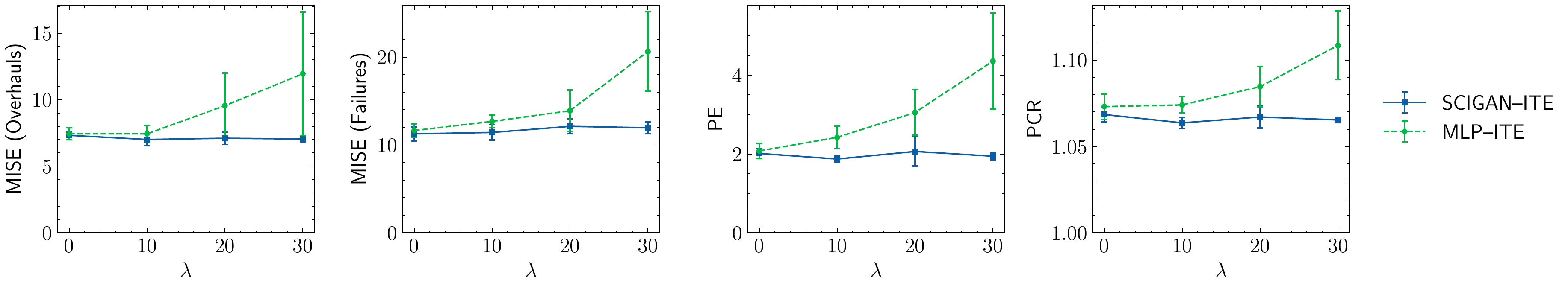}
    \caption{\textbf{Results for varying levels of selection bias.} We show results for different levels of selection bias in terms of $\lambda$ (\cref{eq:sim_selection_bias_lambda}). Although SCIGAN--ITE performs similar to a MLP--ITE for lower values of $\lambda$, it has better performance for stronger levels of bias in terms of MISE, PE, and PCR. 
    }
    \label{fig:selection_bias_results}
\end{figure*}

\section{Conclusion}\label{sec:Conclusion}

This work proposes a novel prescriptive maintenance policy that accounts for imperfect maintenance effects by learning a machine-dependent maintenance effect conditional on the machine's characteristics from observational data. This is achieved by relying on state-of-the-art machine learning methodologies for causal inference. The benefit of our approach is that, unlike existing approaches, our methodology does need strong assumptions regarding the maintenance effect, but is instead able to learn it from observational data using flexible machine learning models. We validate our approach with semi-synthetic experiments using real-life data on more than 4,000 full-service maintenance contracts. We find that our proposed approach outperformed both a supervised approach and non-individualized approach in terms of both accuracy and cost of the prescribed preventive maintenance schedules. Moreover, our work highlights the importance of dealing with selection bias when learning from observational data. These findings show that our proposed approach offers a powerful and flexible policy for individualized maintenance.

Causal inference requires strong assumptions, as does the proposed methodology proposed in this work. The first, overlap, implies overlap between distributions of machine's receiving different levels of maintenance. Overlap can be tested \citep{lei2021distribution} and characterized \citep{oberst2020characterization} from data. Moreover, recent work has looked at characterizing uncertainty in regions where overlap is violated \citep{nethery2019estimating, jesson2020identifying}. The second assumption, unconfoundedness, is untestable in practice \citep{imbens2000role}. It can however be assessed by people with domain-knowledge that are in charge of making maintenance decisions. The relevant question is whether all relevant variables regarding past maintenance decisions are included in the data. If there are unobserved confounders, adequately adjusting for selection bias might not be possible, which would result in biased estimates of the potential outcomes. Recent work has suggested the possibility of sensitivity analyses to assess the effect of hidden confounders \citep{damour2019unobserved, franks2019flexible}. Finally, quantifying ignorance regarding the potential outcomes due to possible violations of these assumptions has been proposed \citep{jesson2021quantifying}.

In terms of future work, it would be valuable to consider different types of maintenance interventions in terms of intensities and costs. Similarly, it would be useful to include more complex costs in this framework, such as stochastic costs or costs that need to be predicted from maintenance or machine characteristics. Moreover, it would be interesting to incorporate more flexible timing of maintenance interventions in our methodology and consider sequences of different maintenance interventions. Sequences of treatments have also received attention in the literature on causal inference \citep[e.g., ][]{robins1999association, hernan2001marginal, bica2019sequences}. Finally, it would be interesting to look at ways of more closely integrating the predictive model in the decision-making step, e.g. by using predict-and-optimize \citep{elmachtoub2022smart} or cost-sensitive approaches \citep{vanderschueren2022predict}.

\section*{Acknowledgements}
This work was supported by the BNP Paribas Fortis Chair in Fraud Analytics, FWO research project G015020N, and FWO PhD Fellowship 11I7322N. The computational resources and services used in this work were provided by the VSC (Flemish Supercomputer Center), funded by the Research Foundation -- Flanders (FWO) and the Flemish Government -- department EWI.

\bibliographystyle{unsrtnat}
\bibliography{references}

\begin{thebibliography}{50}
\providecommand{\natexlab}[1]{#1}
\providecommand{\url}[1]{\texttt{#1}}
\expandafter\ifx\csname urlstyle\endcsname\relax
  \providecommand{\doi}[1]{doi: #1}\else
  \providecommand{\doi}{doi: \begingroup \urlstyle{rm}\Url}\fi

\bibitem[Pham and Wang(1996)]{pham1996imperfect}
Hoang Pham and Hongzhou Wang.
\newblock Imperfect maintenance.
\newblock \emph{European Journal of Operational Research}, 94\penalty0
  (3):\penalty0 425--438, 1996.

\bibitem[Wang(2002)]{wang2002survey}
Hongzhou Wang.
\newblock A survey of maintenance policies of deteriorating systems.
\newblock \emph{European Journal of Operational Research}, 139\penalty0
  (3):\penalty0 469--489, 2002.

\bibitem[Ding and Kamaruddin(2015)]{ding2015maintenance}
Siew-Hong Ding and Shahrul Kamaruddin.
\newblock Maintenance policy optimization—literature review and directions.
\newblock \emph{The International Journal of Advanced Manufacturing
  Technology}, 76\penalty0 (5):\penalty0 1263--1283, 2015.

\bibitem[de~Jonge and Scarf(2020)]{de2020review}
Bram de~Jonge and Philip~A Scarf.
\newblock A review on maintenance optimization.
\newblock \emph{European Journal of Operational Research}, 285\penalty0
  (3):\penalty0 805--824, 2020.

\bibitem[Chukova et~al.(2004)Chukova, Arnold, and Wang]{chukova2004warranty}
S~Chukova, R~Arnold, and Dong~Q Wang.
\newblock Warranty analysis: An approach to modeling imperfect repairs.
\newblock \emph{International journal of production economics}, 89\penalty0
  (1):\penalty0 57--68, 2004.

\bibitem[Nakagawa(1979{\natexlab{a}})]{nakagawa1979imperfect}
Toshio Nakagawa.
\newblock Imperfect preventive-maintenance.
\newblock \emph{IEEE Transactions on Reliability}, 28\penalty0 (5):\penalty0
  402--402, 1979{\natexlab{a}}.

\bibitem[Nakagawa(1979{\natexlab{b}})]{nakagawa1979optimum}
Toshio Nakagawa.
\newblock Optimum policies when preventive maintenance is imperfect.
\newblock \emph{IEEE Transactions on Reliability}, 28\penalty0 (4):\penalty0
  331--332, 1979{\natexlab{b}}.

\bibitem[Brown and Proschan(1983)]{brown1983imperfect}
Mark Brown and Frank Proschan.
\newblock Imperfect repair.
\newblock \emph{Journal of Applied Probability}, 20\penalty0 (4):\penalty0
  851--859, 1983.

\bibitem[Block et~al.(1985)Block, Borges, and Savits]{block1985age}
Henry~W Block, Wagner~S Borges, and Thomas~H Savits.
\newblock Age-dependent minimal repair.
\newblock \emph{Journal of Applied Probability}, 22\penalty0 (2):\penalty0
  370--385, 1985.

\bibitem[Malik(1979)]{malik1979reliable}
Mazhar Ali~Khan Malik.
\newblock Reliable preventive maintenance scheduling.
\newblock \emph{AIIE Transactions}, 11\penalty0 (3):\penalty0 221--228, 1979.

\bibitem[Kijima(1989)]{kijima1989some}
Masaaki Kijima.
\newblock Some results for repairable systems with general repair.
\newblock \emph{Journal of Applied Probability}, 26\penalty0 (1):\penalty0
  89--102, 1989.

\bibitem[Tanwar et~al.(2014)Tanwar, Rai, and Bolia]{tanwar2014imperfect}
Monika Tanwar, Rajiv~N Rai, and Nomesh Bolia.
\newblock Imperfect repair modeling using kijima type generalized renewal
  process.
\newblock \emph{Reliability Engineering \& System Safety}, 124:\penalty0
  24--31, 2014.

\bibitem[Bousdekis et~al.(2021)Bousdekis, Lepenioti, Apostolou, and
  Mentzas]{bousdekis2021review}
Alexandros Bousdekis, Katerina Lepenioti, Dimitris Apostolou, and Gregoris
  Mentzas.
\newblock A review of data-driven decision-making methods for industry 4.0
  maintenance applications.
\newblock \emph{Electronics}, 10\penalty0 (7):\penalty0 828, 2021.

\bibitem[Gits(1992)]{gits1992design}
CW~Gits.
\newblock Design of maintenance concepts.
\newblock \emph{International journal of production economics}, 24\penalty0
  (3):\penalty0 217--226, 1992.

\bibitem[Alaswad and Xiang(2017)]{alaswad2017review}
Suzan Alaswad and Yisha Xiang.
\newblock A review on condition-based maintenance optimization models for
  stochastically deteriorating system.
\newblock \emph{Reliability engineering \& system safety}, 157:\penalty0
  54--63, 2017.

\bibitem[Swanson(2001)]{swanson2001linking}
Laura Swanson.
\newblock Linking maintenance strategies to performance.
\newblock \emph{International journal of production economics}, 70\penalty0
  (3):\penalty0 237--244, 2001.

\bibitem[Carvalho et~al.(2019)Carvalho, Soares, Vita, Francisco, Basto, and
  Alcal{\'a}]{carvalho2019systematic}
Thyago~P Carvalho, Fabr{\'\i}zzio~AAMN Soares, Roberto Vita, Roberto da~P
  Francisco, Jo{\~a}o~P Basto, and Symone~GS Alcal{\'a}.
\newblock A systematic literature review of machine learning methods applied to
  predictive maintenance.
\newblock \emph{Computers \& Industrial Engineering}, 137:\penalty0 106024,
  2019.

\bibitem[Rubin(1974)]{rubin1974estimating}
Donald~B Rubin.
\newblock Estimating causal effects of treatments in randomized and
  nonrandomized studies.
\newblock \emph{Journal of educational Psychology}, 66\penalty0 (5):\penalty0
  688, 1974.

\bibitem[Yao et~al.(2021)Yao, Chu, Li, Li, Gao, and Zhang]{yao2021survey}
Liuyi Yao, Zhixuan Chu, Sheng Li, Yaliang Li, Jing Gao, and Aidong Zhang.
\newblock A survey on causal inference.
\newblock \emph{ACM Transactions on Knowledge Discovery from Data (TKDD)},
  15\penalty0 (5):\penalty0 1--46, 2021.

\bibitem[Imbens(2000)]{imbens2000role}
Guido~W Imbens.
\newblock The role of the propensity score in estimating dose-response
  functions.
\newblock \emph{Biometrika}, 87\penalty0 (3):\penalty0 706--710, 2000.

\bibitem[Hirano and Imbens(2004)]{hirano2004propensity}
Keisuke Hirano and Guido~W Imbens.
\newblock The propensity score with continuous treatments.
\newblock \emph{Applied Bayesian modeling and causal inference from
  incomplete-data perspectives}, 226164:\penalty0 73--84, 2004.

\bibitem[Imai and Van~Dyk(2004)]{imai2004causal}
Kosuke Imai and David~A Van~Dyk.
\newblock Causal inference with general treatment regimes: Generalizing the
  propensity score.
\newblock \emph{Journal of the American Statistical Association}, 99\penalty0
  (467):\penalty0 854--866, 2004.

\bibitem[Schwab et~al.(2020)Schwab, Linhardt, Bauer, Buhmann, and
  Karlen]{schwab2020learning}
Patrick Schwab, Lorenz Linhardt, Stefan Bauer, Joachim~M Buhmann, and Walter
  Karlen.
\newblock Learning counterfactual representations for estimating individual
  dose-response curves.
\newblock In \emph{Proceedings of the AAAI Conference on Artificial
  Intelligence}, volume~34, pages 5612--5619, 2020.

\bibitem[Bica et~al.(2020)Bica, Jordon, and van~der Schaar]{bica2020estimating}
Ioana Bica, James Jordon, and Mihaela van~der Schaar.
\newblock Estimating the effects of continuous-valued interventions using
  generative adversarial networks.
\newblock \emph{Advances in Neural Information Processing Systems},
  33:\penalty0 16434--16445, 2020.

\bibitem[Berrevoets et~al.(2020)Berrevoets, Jordon, Bica, van~der Schaar,
  et~al.]{berrevoets2020organite}
Jeroen Berrevoets, James Jordon, Ioana Bica, Mihaela van~der Schaar, et~al.
\newblock Organite: Optimal transplant donor organ offering using an individual
  treatment effect.
\newblock \emph{Advances in Neural Information Processing Systems}, 33, 2020.

\bibitem[Athey and Wager(2021)]{athey2021policy}
Susan Athey and Stefan Wager.
\newblock Policy learning with observational data.
\newblock \emph{Econometrica}, 89\penalty0 (1):\penalty0 133--161, 2021.

\bibitem[Varian(2016)]{varian2016causal}
Hal~R Varian.
\newblock Causal inference in economics and marketing.
\newblock \emph{Proceedings of the National Academy of Sciences}, 113\penalty0
  (27):\penalty0 7310--7315, 2016.

\bibitem[Devriendt et~al.(2018)Devriendt, Moldovan, and
  Verbeke]{devriendt2018literature}
Floris Devriendt, Darie Moldovan, and Wouter Verbeke.
\newblock A literature survey and experimental evaluation of the
  state-of-the-art in uplift modeling: A stepping stone toward the development
  of prescriptive analytics.
\newblock \emph{Big Data}, 6\penalty0 (1):\penalty0 13--41, 2018.

\bibitem[Verbeke et~al.(2020)Verbeke, Olaya, Berrevoets, Verboven, and
  Maldonado]{verbeke2020foundations}
Wouter Verbeke, Diego Olaya, Jeroen Berrevoets, Sam Verboven, and Sebasti{\'a}n
  Maldonado.
\newblock The foundations of cost-sensitive causal classification.
\newblock \emph{arXiv preprint arXiv:2007.12582}, 2020.

\bibitem[Verbeke et~al.(2022)Verbeke, Olaya, Guerry, and
  Van~Belle]{verbeke2022or}
Wouter Verbeke, Diego Olaya, Marie-Anne Guerry, and Jente Van~Belle.
\newblock To do or not to do? cost-sensitive causal classification with
  individual treatment effect estimates.
\newblock \emph{European Journal of Operational Research}, 2022.

\bibitem[Bertsimas et~al.(2019)Bertsimas, Dunn, and
  Mundru]{bertsimas2019optimal}
Dimitris Bertsimas, Jack Dunn, and Nishanth Mundru.
\newblock Optimal prescriptive trees.
\newblock \emph{INFORMS Journal on Optimization}, 1\penalty0 (2):\penalty0
  164--183, 2019.

\bibitem[Bertsimas and Kallus(2020)]{bertsimas2020predictive}
Dimitris Bertsimas and Nathan Kallus.
\newblock From predictive to prescriptive analytics.
\newblock \emph{Management Science}, 66\penalty0 (3):\penalty0 1025--1044,
  2020.

\bibitem[Deprez et~al.(2021)Deprez, Antonio, and Boute]{deprez2021pricing}
Laurens Deprez, Katrien Antonio, and Robert Boute.
\newblock Pricing service maintenance contracts using predictive analytics.
\newblock \emph{European Journal of Operational Research}, 290\penalty0
  (2):\penalty0 530--545, 2021.

\bibitem[Rubin(2004)]{rubin2004direct}
Donald~B Rubin.
\newblock Direct and indirect causal effects via potential outcomes.
\newblock \emph{Scandinavian Journal of Statistics}, 31\penalty0 (2):\penalty0
  161--170, 2004.

\bibitem[Rubin(2005)]{rubin2005causal}
Donald~B Rubin.
\newblock Causal inference using potential outcomes: Design, modeling,
  decisions.
\newblock \emph{Journal of the American Statistical Association}, 100\penalty0
  (469):\penalty0 322--331, 2005.

\bibitem[Faccio et~al.(2014)Faccio, Persona, Sgarbossa, and
  Zanin]{faccio2014industrial}
M~Faccio, A~Persona, F~Sgarbossa, and G~Zanin.
\newblock Industrial maintenance policy development: A quantitative framework.
\newblock \emph{International Journal of Production Economics}, 147:\penalty0
  85--93, 2014.

\bibitem[Holland(1986)]{holland1986statistics}
Paul~W Holland.
\newblock Statistics and causal inference.
\newblock \emph{Journal of the American Statistical Association}, 81\penalty0
  (396):\penalty0 945--960, 1986.

\bibitem[Silva(2016)]{silva2016observational}
Ricardo Silva.
\newblock Observational-interventional priors for dose-response learning.
\newblock \emph{Advances in Neural Information Processing Systems}, 29, 2016.

\bibitem[Lei et~al.(2021)Lei, D’Amour, Ding, Feller, and
  Sekhon]{lei2021distribution}
Lihua Lei, Alexander D’Amour, Peng Ding, Avi Feller, and Jasjeet Sekhon.
\newblock Distribution-free assessment of population overlap in observational
  studies.
\newblock 2021.

\bibitem[Oberst et~al.(2020)Oberst, Johansson, Wei, Gao, Brat, Sontag, and
  Varshney]{oberst2020characterization}
Michael Oberst, Fredrik Johansson, Dennis Wei, Tian Gao, Gabriel Brat, David
  Sontag, and Kush Varshney.
\newblock Characterization of overlap in observational studies.
\newblock In \emph{International Conference on Artificial Intelligence and
  Statistics}, pages 788--798. PMLR, 2020.

\bibitem[Nethery et~al.(2019)Nethery, Mealli, and
  Dominici]{nethery2019estimating}
Rachel~C Nethery, Fabrizia Mealli, and Francesca Dominici.
\newblock Estimating population average causal effects in the presence of
  non-overlap: The effect of natural gas compressor station exposure on cancer
  mortality.
\newblock \emph{The annals of applied statistics}, 13\penalty0 (2):\penalty0
  1242, 2019.

\bibitem[Jesson et~al.(2020)Jesson, Mindermann, Shalit, and
  Gal]{jesson2020identifying}
Andrew Jesson, S{\"o}ren Mindermann, Uri Shalit, and Yarin Gal.
\newblock Identifying causal-effect inference failure with uncertainty-aware
  models.
\newblock \emph{Advances in Neural Information Processing Systems},
  33:\penalty0 11637--11649, 2020.

\bibitem[D'Amour(2019)]{damour2019unobserved}
Alexander D'Amour.
\newblock On multi-cause approaches to causal inference with unobserved
  counfounding: Two cautionary failure cases and a promising alternative.
\newblock In Kamalika Chaudhuri and Masashi Sugiyama, editors,
  \emph{Proceedings of the Twenty-Second International Conference on Artificial
  Intelligence and Statistics}, volume~89 of \emph{Proceedings of Machine
  Learning Research}, pages 3478--3486. PMLR, 16--18 Apr 2019.
\newblock URL \url{https://proceedings.mlr.press/v89/d-amour19a.html}.

\bibitem[Franks et~al.(2019)Franks, D’Amour, and Feller]{franks2019flexible}
AlexanderM Franks, Alexander D’Amour, and Avi Feller.
\newblock Flexible sensitivity analysis for observational studies without
  observable implications.
\newblock \emph{Journal of the American Statistical Association}, 2019.

\bibitem[Jesson et~al.(2021)Jesson, Mindermann, Gal, and
  Shalit]{jesson2021quantifying}
Andrew Jesson, S{\"o}ren Mindermann, Yarin Gal, and Uri Shalit.
\newblock Quantifying ignorance in individual-level causal-effect estimates
  under hidden confounding.
\newblock In \emph{International Conference on Machine Learning}, pages
  4829--4838. PMLR, 2021.

\bibitem[Robins(1999)]{robins1999association}
James~M Robins.
\newblock Association, causation, and marginal structural models.
\newblock \emph{Synthese}, pages 151--179, 1999.

\bibitem[Hern{\'a}n et~al.(2001)Hern{\'a}n, Brumback, and
  Robins]{hernan2001marginal}
Miguel~A Hern{\'a}n, Babette Brumback, and James~M Robins.
\newblock Marginal structural models to estimate the joint causal effect of
  nonrandomized treatments.
\newblock \emph{Journal of the American Statistical Association}, 96\penalty0
  (454):\penalty0 440--448, 2001.

\bibitem[Bica et~al.(2019)Bica, Alaa, Jordon, and van~der
  Schaar]{bica2019sequences}
Ioana Bica, Ahmed~M Alaa, James Jordon, and Mihaela van~der Schaar.
\newblock Estimating counterfactual treatment outcomes over time through
  adversarially balanced representations.
\newblock In \emph{International Conference on Learning Representations}, 2019.

\bibitem[Elmachtoub and Grigas(2022)]{elmachtoub2022smart}
Adam~N Elmachtoub and Paul Grigas.
\newblock Smart “predict, then optimize”.
\newblock \emph{Management Science}, 68\penalty0 (1):\penalty0 9--26, 2022.

\bibitem[Vanderschueren et~al.(2022)Vanderschueren, Verdonck, Baesens, and
  Verbeke]{vanderschueren2022predict}
Toon Vanderschueren, Tim Verdonck, Bart Baesens, and Wouter Verbeke.
\newblock Predict-then-optimize or predict-and-optimize? an empirical
  evaluation of cost-sensitive learning strategies.
\newblock \emph{Information Sciences}, 594:\penalty0 400--415, 2022.

\end{thebibliography}

\end{document}